\documentclass{PoS}
\usepackage{epsf,amsmath,amssymb,axodraw,graphicx,scalefnt,subfigure}

\newcommand{\eft}{{\abbrev EFT}}

\newcommand{\qcd}{{\abbrev QCD}}

\newcommand{\abbrev}{\scalefont{.9}\rm}

\newcommand{\mhiggs}{M_{\rm H}}
\newcommand{\mtop}{M_{t}}

\newcommand{\eqn}[1]{Eq.\,(\ref{#1})}
\newcommand{\fig}[1]{Fig.\,\ref{#1}}

\newcommand{\order}[1]{{\cal O}(#1)}
\newcommand{\lhc}{{\abbrev LHC}}

\newcommand{\lo}{{\abbrev LO}}
\newcommand{\nlo}{{\abbrev NLO}}
\newcommand{\nnlo}{{\abbrev NNLO}}

\title{Higgs production in gluon fusion at {\scalefont{.9}NNLO} for finite
  top quark mass\thanks{Work supported by {\abbrev DFG} under contract
HA~2990/3-1, and the Helmholtz Alliance ``Physics at the Terascale''.}}

\ShortTitle{Higgs production at finite top mass}

\author{\speaker{Robert Harlander}\\
        Fachbereich C, Bergische Universit\"at
        Wuppertal, 42097 Wuppertal, Germany\\
        E-mail: \email{robert.harlander@uni-wuppertal.de}}
\author{Hendrik Mantler\\
        Fachbereich C, Bergische Universit\"at
        Wuppertal, 42097 Wuppertal, Germany\\
        E-mail: \email{	hendrik.mantler@uni-wuppertal.de}}

\author{Simone Marzani\\
  School of Physics \& Astronomy, University of Manchester, Manchester
  M13 9PL, UK\\
  E-mail: \email{ simone.marzani@manchester.ac.uk }}
\author{Kemal Ozeren\\
        Fachbereich C, Bergische Universit\"at
        Wuppertal, 42097 Wuppertal, Germany\\
        E-mail: \email{ozeren@physik.uni-wuppertal.de}}

\abstract{ The evaluation of the top quark mass suppressed terms to the
  Higgs production cross section in gluon fusion at
  next-to-next-to-leading order is reported on. In the region below
  threshold, the Feynman diagrams are evaluated using asymptotic
  expansions. The result is then matched to the high-energy limit
  derived from $k_T$ factorization. The result shows that the heavy-top
  limit used so far approximates the full result to better than 1\%.}

\FullConference{RADCOR 2009 - 9th International Symposium on Radiative Corrections (Applications of Quantum Field Theory to Phenomenology) ,\\
		 October 25 - 30 2009\\
		 Ascona, Switzerland}

\begin{document}

\section{Introduction}

The importance of the gluon fusion mechanism for Higgs production at
hadron colliders has been highlighted in the recent past by the first
statistically significant exclusion limits of the combined CDF/D0
searches for the Higgs boson at the Tevatron
collider~\cite{Collaboration:2009je}. This result depends crucially on
the knowledge of higher order radiative corrections. If only the leading
order result for the gluon fusion cross section had been taken into
account in the analyses, for example, a 95\% exclusion would be way out
of reach, even for years. The \nlo{} results increase the theoretical
prediction by more than 100\%~\cite{Dawson:1990zj,Djouadi:1991tk}. Yet,
the sensitivity would still be insufficient to claim exclusion. It is
only the \nnlo{} result that allows such a claim.

In this light, it is important to ensure the validity of the theoretical
prediction. The \nnlo{} \qcd{} result that goes into the experimental
analyses has been evaluated by (more than) three different
groups~\cite{Harlander:2002wh,Anastasiou:2002yz,Ravindran:2003um}. Also,
various studies based on resummation have convincingly shown that we do
not have to expect crucially large numerical contributions from \qcd{}
beyond \nnlo{} (see, e.g., Refs.\,\cite{Catani:2003zt,Ahrens:2008nc}).

Concerning the electro-weak corrections, they are found to be below 6\%
in the relevant Higgs mass range~\cite{Actis:2008ug}. Unfortunately,
threshold effects from virtual $W$ and $Z$ bosons lead to spurious
spikes in the range 160-190~GeV which are smoothened by the finite
widths of the gauge bosons. Since there is a certain amount of freedom
in this procedure (as in any other combination of all-order and
fixed-order expressions), it is not completely clear to what extent this
reflects in the theoretical uncertainty of this result. Knowing that the
pure \qcd{} corrections are large, one may expect that \qcd{} effects
further enhance this uncertainty. In fact, an explicit calculation of
the mixed electro-weak/\qcd{} effects (albeit in the limit $M_H\ll M_W$)
confirms this~\cite{Anastasiou:2008tj}.

Further worries may concern the use of the inclusive ($M_H$ dependent)
K-factor in the experimental analysis. However, fully exclusive \nnlo{}
calculations for gluon fusion are available and can be used to check the
efficiencies~\cite{Anastasiou:2009bt}.

In this proceedings contribution, we will report on works that have
addressed another issue which has plagued the \nnlo{} results mentioned
before, namely the effects arising from a finite top quark mass.

\section{Effective field theory approach}

Due to their high complexity, calculations of the gluon fusion process
beyond the inclusive \nlo{} cross section were all performed in the
so-called {\it effective field theory (\eft{}) approach}. This means
that the six-flavor Lagrangian is replaced by
\begin{equation}
\begin{split}
{\cal L}_{\rm eff} = -\frac{1}{4v}C_1\,H\,G_{\mu\nu}G^{\mu\nu} + {\cal
  L}^{(5)}_{\qcd}\,,
\label{eq::leff}
\end{split}
\end{equation}
where ${\cal L}_{\qcd}^{(5)}$ is the five-flavor \qcd{} Lagrangian (no
top-quark), and $G_{\mu\nu}$ is the \qcd{} field strength tensor. The
Wilson coefficient $C_1$ is known to {\abbrev
  N$^4$LO}~\cite{Schroder:2005hy,Chetyrkin:2005ia}.  In the \eft{}
approach, the loop-induced gluon-Higgs coupling is thus replaced by a
tree-level coupling proportional to $C_1$.

Clearly, a result derived from \eqn{eq::leff} cannot be expected to hold
beyond the top quark threshold, $\mhiggs > 2\mtop$. However, at \nlo{}
one observes that the bulk of the top quark mass dependence is given by
the \lo{} cross section. Therefore, whenever we speak of the \eft{}
approach in this paper, we mean the expression
\begin{equation}
\begin{split}
\sigma_\infty \equiv
\sigma^{(0)}(\mtop)\frac{\sigma(\mtop\to\infty)}{\sigma^{(0)}(\mtop\to\infty)}\,,
\label{eq::eft}
\end{split}
\end{equation}
where $\sigma^{(0)}$ is the leading order term in $\alpha_s$.  Even
though the exact \nlo{} result is approximated in the \eft{} approach to
better than 1\% below threshold, its validity at \nnlo{} remained a
matter of concern. At first sight, an obvious way to check it is to
calculate top quark mass suppressed terms to the total cross section and
ensure that they do not significantly alter the \eft{} result. The next
section describes the corresponding calculations.

\section{Top quark mass suppressed terms}\label{sec::mtsup}

Due to the absence of a gluon-Higgs vertex in the Standard Model, there
is no tree-level contribution to $\sigma(gg\to H+X)$. All the
corresponding Feynman diagrams contain a closed quark loop that mediates
this coupling. The dominant contribution is due to a top quark loop; the
bottom loop contribution amounts to only a few percent at \lo{}.  In the
\eft{} approach, the top quark is integrated out, resulting in a direct
gluon-Higgs coupling as described by the Lagrangian in \eqn{eq::leff},
and the number of loops in the Feynman diagrams reduces by one.

Alternatively to the \eft{} approach, one can evaluate the Feynman
diagrams approximately with the help of the method of {\it asymptotic
  expansions}~(see, e.g., Ref.\,\cite{Smirnov:1994tg}). They allow one
to obtain a systematic expansion of the relevant partonic cross sections
in terms of powers and logarithms of $\mhiggs^2/\mtop^2$. The first term
in this expansion will then agree with the result obtained from
\eqn{eq::leff}, but higher orders can be obtained in a straightforward
manner by increasing the depth of intermediate Taylor expansions.

The method of asymptotic expansions expresses the Feynman diagrams under
consideration in terms of products of massive vacuum and massless vertex
or box integrals. The former ones are required through three loops and
can be evaluated with the help of the {\tt
  FORM}~\cite{Vermaseren:2000nd} program {\tt
  MATAD}~\cite{Steinhauser:2000ry}. The vertex integrals are needed
through two loops: in Ref.\,\cite{Harlander:2009bw,Harlander:2009mq}, they were
calculated using the method of Ref.\,\cite{Baikov:2000jg} as implemented
in Ref.\,\cite{Harlander:2000mg} (the implementation is based on the
{\tt FORM} version of {\tt MINCER}~\cite{Larin:1991fz}).  The massless
boxes are only needed at the one-loop level and can be calculated by
standard methods.

Note that the massless component of the $2\to 3$ processes is given by
tree-level diagrams. However, this class is the most difficult one as
far as the phase space integration is concerned. 
In Ref.\,\cite{Harlander:2009mq}, these integrals were evaluated in
terms of expansions around $\hat s=\mhiggs^2$. As will be explained
below, this approximation is fully justified in the the approach applied
here.

Explicit results for all the partonic cross sections have been presented
in Refs.\,\cite{Harlander:2009bw,Harlander:2009mq,
  Pak:2009bx,Pak:2009dg}.\footnote{The virtual terms were obtained
  through $\order{1/\mtop^6}$ and $\order{1/\mtop^8}$ in
  Ref.~\cite{Harlander:2009bw} and Ref.\,\cite{Pak:2009bx},
  respectively, while the real radiation contributions were obtained
  through $\order{1/\mtop^6}$ in Ref.~\cite{Harlander:2009mq} and
  $\order{1/\mtop^4}$ in Ref.~\cite{Pak:2009dg}.}

\section{Large-$\hat s$ region}\label{sec::larges}
The expansion described in Section~\ref{sec::mtsup} is obtained by
assuming that the top quark mass is the largest mass scale of the
physical system. This is, of course, not true in reality, because $\hat
s$, the partonic center-of-mass energy, assumes values up to the
hadronic center-of-mass energy $s$ (i.e., $1.96$~TeV at the Tevatron,
and -- hopefully -- $14$~TeV at the \lhc).  In fact, this very same
issue arises already in the \eft{} approach. Fortunately, however, the
parton luminosity becomes very small at large $\hat s$. In fact, at
\nlo{} one observes that the hadronic cross section is approximated to
better than 90\% by neglecting contributions from $\sqrt{\hat s}> 2\mtop$.

Including higher orders in the $1/\mtop$ expansion, however, the problem
becomes more severe. The reason is that the expansion of
Section~\ref{sec::mtsup} generates terms of the form $(\hat
s/\mtop^2)^k$, leading to a power divergence at large $\hat s$.  By
coincidence, at \nlo{} the coefficient of the $k=1$ term vanishes (for
the $gg$ initial state). This observation was used in
Ref.\,\cite{Dawson:1993qf} to derive an estimate of the top mass
suppressed terms at \nlo{}.

The failure of the $1/\mtop$ expansion for $\sqrt{\hat s}>2\mtop$ is
also the reason why the so-called soft expansion around $\hat
s=\mhiggs^2$ for the phase space integrals mentioned in
Section~\ref{sec::mtsup} is fully sufficient: within $\mhiggs^2<\hat
s<4\mtop^2$, it is expected (and observed) to converge well, while
outside this region, the $1/\mtop$ expansion breaks down anyway.

At \nnlo{}, we see no reason why the $\hat s/\mtop^2$ term should vanish
as well. In addition, the goal of
Refs.\,\cite{Harlander:2009mq,Harlander:2009my} was to derive not only
an estimate of the top mass effects, but to provide a consistent
quantitative approximation of these terms. This could be achieved from
an additional piece of information which had recently been
evaluated~\cite{Marzani:2008az}, name the true large-$\hat s$ limit of
the partonic cross sections.

Using this information, an expression for the full partonic cross
section that incorporates all known information on the \nnlo{} cross
section can be constructed as follows:
\begin{equation}
\begin{split}
\hat\sigma^{(n)}_{\alpha\beta}(x) = \hat\sigma^{(n)}_{\alpha\beta,N}(x)
&+ \,\sigma_0 A^{(n)}_{\alpha\beta}\left[ \ln \frac{1}{x} -
  \sum_{k=1}^N\frac{1}{k}(1-x)^k  \right]
+ (1-x)^{N+1}\,\left[\sigma_0 B_{\alpha\beta}^{(n)} -
  \hat\sigma^{(n)}_{\alpha\beta,N}(0)
  \right]\,,
\label{eq::match}
\end{split}
\end{equation}
where $\hat\sigma^{(n)}_{\alpha\beta,N}(x)$ denotes the soft expansion
of the {\abbrev N}$^n${\abbrev LO} partonic cross section for the
process $\alpha\beta\to H+X$ through order $(1-x)^{N}$, where
$x=\mhiggs^2/\hat s$. The coefficients $A_{\alpha\beta}^{(n)}$ and
$B_{\alpha\beta}^{(n)}$ determine the behaviour of the partonic cross
section as $x\to 0$. The leading terms at \nlo{} and
\nnlo{} (i.e., $A_{\alpha\beta}^{(1)}=0$, $A_{\alpha\beta}^{(2)}$, and
$B_{\alpha\beta}^{(1)}$) for the $gg$ channel were given in the form of
numerical tables in Ref.\,\cite{Marzani:2008az}, and for the other
channels in Ref.\,\cite{Harlander:2009my}.

The quality of this approach can be tested at \nlo{} by comparing it to
the exact result which is known in numerical form~(see, e.g.,
Ref.\,\cite{Spira:1995rr}). One observes excellent agreement for the
$gg$ and the $qg$ channel for the relevant Higgs mass range between 100
and 300~GeV, while the $q\bar q$ channel appears to be more
problematic. This is due to the fact that the only diagram contributing
to this channel vanishes at both small and large $x$. This leaves room
for a relatively pronounced structure at threshold which cannot be
described properly by our approach. However, the $q\bar q$ channel is
down by almost three orders of magnitude relative to the $gg$ channel.

At \nnlo{}, the unknown constants $B_{\alpha\beta}^{(2)}$ introduced a
certain amount of uncertainty to the prediction. In
Ref.\,\cite{Harlander:2009my} it was estimated to be of the order of
1\%, where also more detailed studies of the partonic cross sections can
be found.

\section{Hadronic cross section}

\begin{figure}
  \begin{center}
      \includegraphics[bb=110 265 465
        560,width=.45\textwidth]{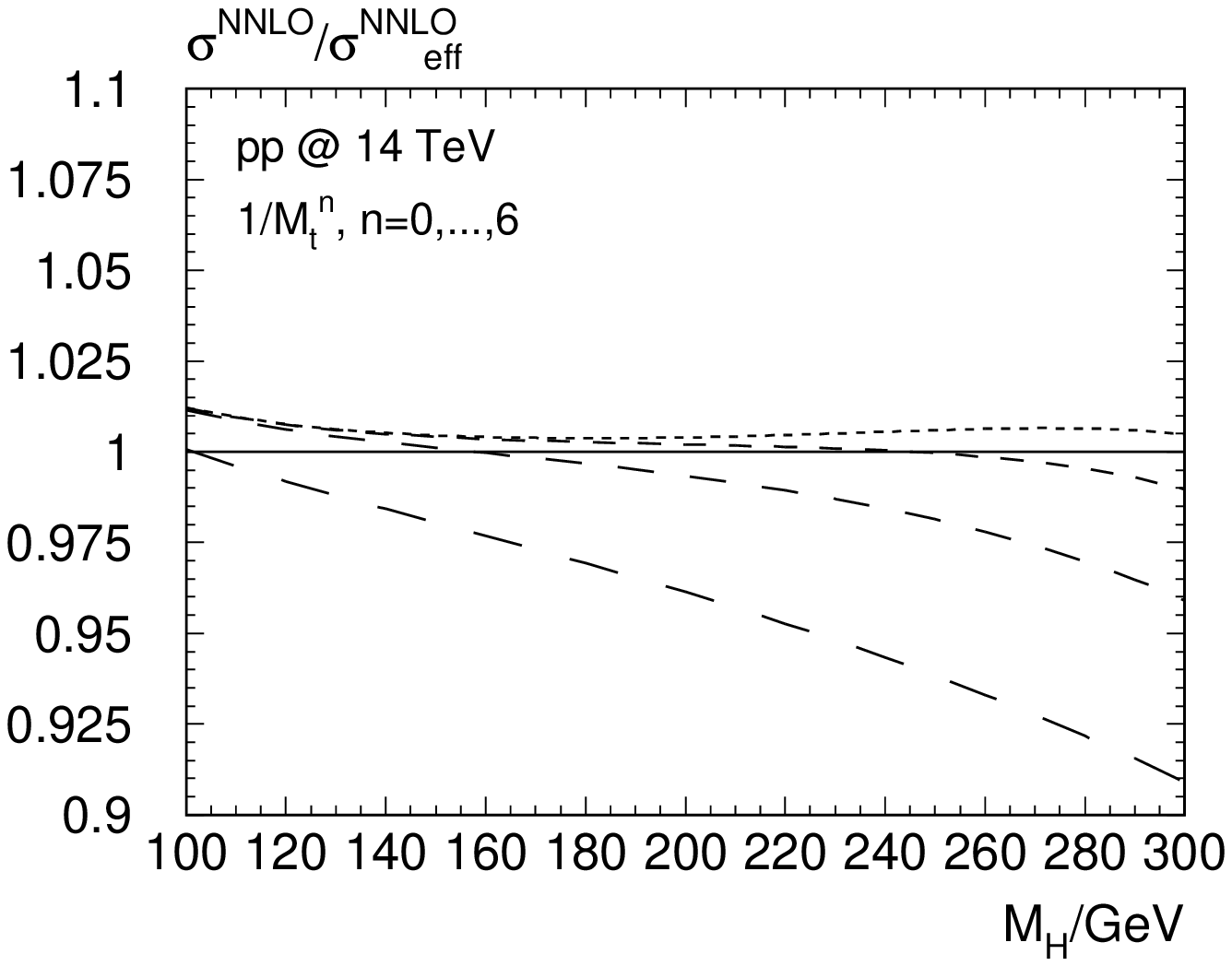}\qquad
    \includegraphics[bb=110 265 465
      560,width=.45\textwidth]{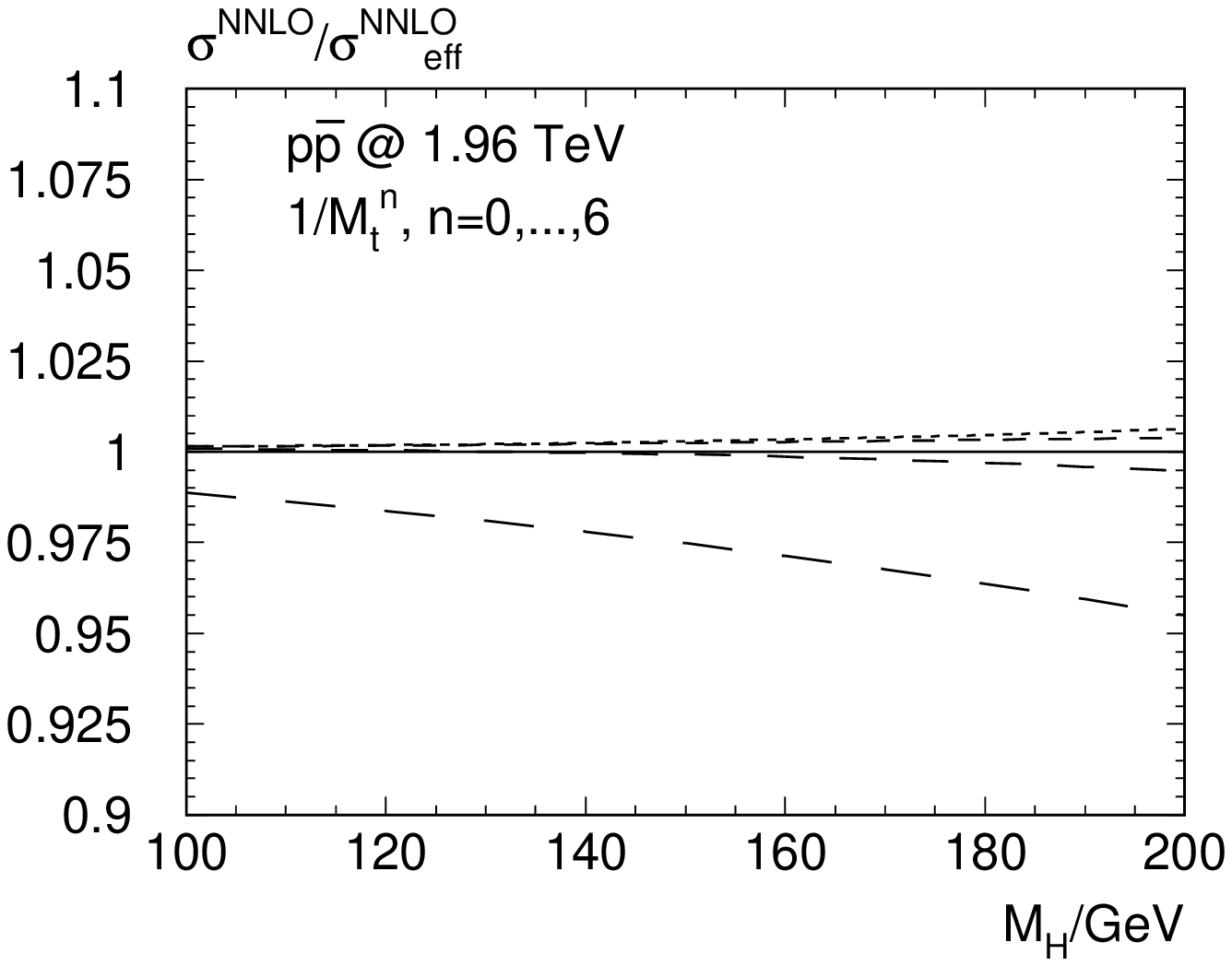}
    \parbox{.9\textwidth}{
      \caption[]{\label{fig::ratnnlo}\sloppy
        Ratio of the hadronic cross section as obtained from
        \eqn{eq::match} to the \eft{} result (see Eqs.\,(\ref{eq::leff})
        and (\ref{eq::eft})). The
        various lines correspond to keeping different orders in
        $1/\mtop$ in the numerator $\sigma^\nnlo{}$ (decreasing
        dash-length corresponds to increasing order in $1/\mtop$).
        From Ref.\,\cite{Harlander:2009my}.
  }}
  \end{center}
\end{figure}

The hadronic cross section is obtained by integrating the partonic
expression from \eqn{eq::match} over the parton densities (we use
{\abbrev MSTW2008}~\cite{Martin:2009iq}). The most important question is
how well the \eft{} approximation (i.e., keeping the leading term in the
$1/\mtop$ expansion and factoring out the full mass dependent result at
\lo{} in $\alpha_s$) describes the top quark mass effects. We therefore
show in \fig{fig::ratnnlo} the ratio of our result to the \eft{} approach,
both for the \lhc{} and the Tevatron. The agreement in both cases is
better than 1\% which is well below the current estimated theoretical
uncertainty due to higher orders in $\alpha_s$ and {\abbrev PDF}
variations.

This is a very comforting result since meanwhile a large number of
theoretical and experimental studies have been performed based on the
\eft{} approach.

\section{Conclusions}

The quality of the heavy-top limit used in numerous studies and
calculations for Higgs production in gluon fusion has been scrutinized
by an explicit calculation of the top mass suppressed terms. The result
was derived from asymptotic expansions of the relevant Feynman diagrams
and the combination with the high-energy limit obtained from $k_T$
factorization. The result justifies the use of the effective theory
approach to a very high degree, at least for the inclusive cross
section. It remains to be seen how this result carries over to less
inclusive quantities or phase space restrictions.

Finally, let us point out that a similar, independent
calculation~\cite{Pak:2009dg,Pak:2009bx} was presented also by A.~Pak at
this conference (see these proceedings).

\def\app#1#2#3{{\it Act.~Phys.~Pol.~}\jref{\bf B #1}{#2}{#3}}
\def\apa#1#2#3{{\it Act.~Phys.~Austr.~}\jref{\bf#1}{#2}{#3}}
\def\annphys#1#2#3{{\it Ann.~Phys.~}\jref{\bf #1}{#2}{#3}}
\def\cmp#1#2#3{{\it Comm.~Math.~Phys.~}\jref{\bf #1}{#2}{#3}}
\def\cpc#1#2#3{{\it Comp.~Phys.~Commun.~}\jref{\bf #1}{#2}{#3}}
\def\epjc#1#2#3{{\it Eur.\ Phys.\ J.\ }\jref{\bf C #1}{#2}{#3}}
\def\fortp#1#2#3{{\it Fortschr.~Phys.~}\jref{\bf#1}{#2}{#3}}
\def\ijmpc#1#2#3{{\it Int.~J.~Mod.~Phys.~}\jref{\bf C #1}{#2}{#3}}
\def\ijmpa#1#2#3{{\it Int.~J.~Mod.~Phys.~}\jref{\bf A #1}{#2}{#3}}
\def\jcp#1#2#3{{\it J.~Comp.~Phys.~}\jref{\bf #1}{#2}{#3}}
\def\jetp#1#2#3{{\it JETP~Lett.~}\jref{\bf #1}{#2}{#3}}
\def\jphysg#1#2#3{{\small\it J.~Phys.~G~}\jref{\bf #1}{#2}{#3}}
\def\jhep#1#2#3{{\small\it JHEP~}\jref{\bf #1}{#2}{#3}}
\def\mpl#1#2#3{{\it Mod.~Phys.~Lett.~}\jref{\bf A #1}{#2}{#3}}
\def\nima#1#2#3{{\it Nucl.~Inst.~Meth.~}\jref{\bf A #1}{#2}{#3}}
\def\npb#1#2#3{{\it Nucl.~Phys.~}\jref{\bf B #1}{#2}{#3}}
\def\nca#1#2#3{{\it Nuovo~Cim.~}\jref{\bf #1A}{#2}{#3}}
\def\plb#1#2#3{{\it Phys.~Lett.~}\jref{\bf B #1}{#2}{#3}}
\def\prc#1#2#3{{\it Phys.~Reports }\jref{\bf #1}{#2}{#3}}
\def\prd#1#2#3{{\it Phys.~Rev.~}\jref{\bf D #1}{#2}{#3}}
\def\pR#1#2#3{{\it Phys.~Rev.~}\jref{\bf #1}{#2}{#3}}
\def\prl#1#2#3{{\it Phys.~Rev.~Lett.~}\jref{\bf #1}{#2}{#3}}
\def\pr#1#2#3{{\it Phys.~Reports }\jref{\bf #1}{#2}{#3}}
\def\ptp#1#2#3{{\it Prog.~Theor.~Phys.~}\jref{\bf #1}{#2}{#3}}
\def\ppnp#1#2#3{{\it Prog.~Part.~Nucl.~Phys.~}\jref{\bf #1}{#2}{#3}}
\def\rmp#1#2#3{{\it Rev.~Mod.~Phys.~}\jref{\bf #1}{#2}{#3}}
\def\sovnp#1#2#3{{\it Sov.~J.~Nucl.~Phys.~}\jref{\bf #1}{#2}{#3}}
\def\sovus#1#2#3{{\it Sov.~Phys.~Usp.~}\jref{\bf #1}{#2}{#3}}
\def\tmf#1#2#3{{\it Teor.~Mat.~Fiz.~}\jref{\bf #1}{#2}{#3}}
\def\tmp#1#2#3{{\it Theor.~Math.~Phys.~}\jref{\bf #1}{#2}{#3}}
\def\yadfiz#1#2#3{{\it Yad.~Fiz.~}\jref{\bf #1}{#2}{#3}}
\def\zpc#1#2#3{{\it Z.~Phys.~}\jref{\bf C #1}{#2}{#3}}
\def\ibid#1#2#3{{ibid.~}\jref{\bf #1}{#2}{#3}}
\def\otherjournal#1#2#3#4{{\it #1}\jref{\bf #2}{#3}{#4}}
\newcommand{\jref}[3]{{\bf #1} (#2) #3}
\newcommand{\hepph}[1]{{\tt hep-ph/#1}}
\newcommand{\mathph}[1]{{\tt math-ph/#1}}
\newcommand{\arxiv}[2]{{\tt arXiv:#1}}
\newcommand{\bibentry}[4]{#1, #3.}

\end{document}